\documentclass[prl,aps,twocolumn,showpacs]{revtex4} 
\usepackage{graphicx}  
\newcommand\bx{{\bf x}} 
\begin{document} 
\title{Quantum critical fluctuations in disordered $d$-wave 
  superconductors} 
\author{Julia S. Meyer${}^a$, Igor V. Gornyi${}^{b,*}$, and Alexander 
  Altland${}^a$}

\affiliation{${}^a$Institut f\"ur Theoretische Physik, Universit\"at zu 
K\"oln, Z\"ulpicher Str.~77, 50937 K\"oln, Germany\\
${}^b$Institut f\"ur Nanotechnologie, Forschungszentrum Karlsruhe, 
76021 Karlsruhe, Germany;\\Institut f\"ur Theorie der Kondensierten 
Materie, Universit\"at Karlsruhe, 76128 Karlsruhe, Germany}

\date{\today}


\begin{abstract} 
 
  Quasiparticles in the cuprates appear to be subject to anomalously
  strong inelastic damping mechanisms. To explain the phenomenon,
  Sachdev and collaborators recently proposed to couple the system to
  a critically fluctuating order parameter mode of either $id_{xy}$-
  or $is$-symmetry. Motivated by the observation that the energies
  relevant for the dynamics of this mode are comparable to the
  scattering rate induced by even moderate impurity concentrations, we
  here generalize the approach to the presence of static disorder. In
  the $id$-case, we find that the coupling to disorder renders the
  order parameter dynamics diffusive but otherwise leaves much of the
  phenomenology observed in the clean case intact. In contrast, the
  interplay of impurity scattering and order parameter fluctuations of
  $is$-symmetry entails the formation of a secondary superconductor
  transition, with a critical temperature exponentially sensitive to
  the disorder concentration.

\end{abstract}

\maketitle

Recent experimental work shows that essential aspects of the
low-temperature behaviour of cuprate superconductors lie beyond the
scope of orthodox BCS-type formulations in terms of a non-interacting
quasiparticle (QP) system. Notably, high resolution angle-resolved
photoemission (ARPES) experiments~\cite{ARPES} indicate a QP
relaxation rate $\kappa \sim |\epsilon|$ proportional to the
excitation energy (for temperatures $T<|\epsilon|$), while the
standard BCS picture predicts $\kappa \sim |\epsilon|^3$~\cite{eps3}.
The much weaker $\sim |\epsilon|$ scaling most likely represents a
correlation effect.

Arguing that a qualitative modification of the $|\epsilon|^3$-law
required coupling of the QP system to some kind of collective mode,
Sachdev and collaborators~\cite{SaC} recently proposed a dynamical
generalization of the BCS order parameter field, viz.
\begin{equation} 
    \label{eq:2} 
      \hat \Delta(\bx,t) = \hat \Delta_{0}(\bx) + i\hat \Delta_{1}(\bx,t),     
\end{equation} 
where $\hat \Delta_0$ is the background field of $d_{x^2-y^2}$ angular
momentum dependence and $i \hat\Delta_{1}\equiv i \hat
\Delta^{s/d_{xy}}$ a time-dependent component of either $s$- or
$d_{xy}$-wave symmetry.

The basic idea behind the proposal (\ref{eq:2}) is that somewhere in
close vicinity to the phase diagram of the system there might be a
quantum critical point towards formation of a stable $i\hat \Delta_1$
order parameter amplitude (a $d+is/d_{xy}$ state.)  Although the
actual transition point may lie outside the accessible parameter
space, its proximity will manifest itself through dynamical order
parameter {\it fluctuations}, as described by (\ref{eq:2}).

The coupling of the collective mode $\hat \Delta_1(t)$ to the
relativistic QPs strongly alters the low-energy phenomenology of the
system~\cite{SaC,KP}.  Specifically, (i) scattering off the
time-dependent order parameter fluctuations enhances the QP relaxation
to a linear rate $\kappa \sim |\epsilon|$.  This amplification
mechanism appears to be insensitive to microscopic details such as the
ultraviolet structure of the QP dispersion relation. In particular,
(ii) no qualitative differences between $s$- and $d_{xy}$-scattering
exist. The coupling of the QPs to $\hat \Delta_1$ leads (iii) to the
formation of long-ranged collective fluctuations which
should~\cite{SaC} be observable by neutron or Raman scattering.
However, (iv) if sufficiently weak, the attractive interaction
mediated by $\hat \Delta_1$ does {\it not} open a QP gap, i.e., it
does not drive the system into a secondary superconductor instability.

It is the purpose of this Letter to explore whether elements of the
phenomenology outlined above survive generalization to the presence of
static disorder. In view of the fact that all features (i)-(iv) rely
on a conspiracy of order parameter fluctuations and the Dirac spectrum
of the QPs -- the latter being thoroughly corrupted by impurity
scattering -- one might expect the answer to be largely negative.
However, as we are going to show below this presumption is overly
conservative: owing to a decoupling of the order parameter from the
impurity scattering vertex ($id$-case) and/or the renormalization of
the polarization operator by infrared-singular diffusion modes
($is$-case), the disordered system, too, turns out to be strongly
affected by fluctuations of $\hat \Delta_1(t)$.
    
To summarize our main findings, (i') notwithstanding the presence of
disorder, the coupling of QPs to a dynamical order parameter of
$id$-{\it symmetry} continues to induce an inelastic relaxation rate
$\kappa_{\rm in}\sim |\epsilon|$.  However, this contribution sits now
superimposed on a structureless elastic background $\kappa_{\rm el}
\sim \tau^{-1}$, where $\tau$ denotes the impurity scattering time.
Further, at low frequencies, $|\epsilon|< \tau^{-1}$, a
logarithmically singular third contribution $\kappa_{\rm sing} \sim
\ln (\tau |\epsilon|)$ to the relaxation rate $\kappa = \kappa_{\rm
  in} + \kappa_{\rm el}+ \kappa_{\rm sing}$ signals the formation of
long-ranged diffusive excitations which eventually (for asymptotically
low excitation energies/temperatures) fully absorb the spectral weight
carried by individual QP states. (ii') Unlike in the clean case, the
system is no longer indifferent to the symmetry of $\hat \Delta_1(t)$.
While in the $id$-case, QPs remain (marginally) stable objects and
(iii') the order parameter continues to exhibit fluctuation behaviour
similar to that found in Refs.~\cite{SaC,KP}, a conspiracy of impurity
scattering and dynamical fluctuations of $is$-{\it symmetry} (iv')
drives the system into a secondary superconductor transition.  More
specifically, at the critical temperature
\begin{equation} 
    \label{eq:8} 
    {\tilde T_c} \sim \omega_0 \exp(-\tilde r/\nu_0), 
\end{equation} 
the system enters a $d+is$ superconducting state and the QPs become
gapped.  Here $\omega_0={\rm min} [\omega_D,1/\tau]$, where $\omega_D$
is a phenomenological parameter setting the maximum (Debye) frequency
of the order parameter fluctuations, $1/\tilde r$ denotes the
amplitude of the order parameter fluctuations, and $\nu_0$ the
residual low-energy density of states (DoS) induced by impurity
scattering.  Notice that through the dependence of the DoS on the
impurity concentration, $\nu_0\propto1/\tau$, the transition
temperature varies exponentially with the disorder strength.  Since
the latter can be deliberately changed by doping, the observability of
a disorder-dependent BCS superconductor transition at low temperatures
appears to impose a relatively straightforward criterion testing the
presence of dynamical $is$-order parameter fluctuations in the system.
 
To prepare our discussion of the interplay order parameter
fluctuations/disorder, let us briefly recapitulate the physics of the
clean system.  Denoting the two-component QP field by $\psi^a$, where
$a=1,\bar 1,2,\bar 2$ enumerates the low-energy nodes in the Brillouin
zone, and the time-dependent amplitude of the order parameter by
$\phi(x)$, the low-energy physics of the system is described by a
composite action $S[\phi,\psi] = S[\psi] + S[\phi] + S_{\rm
  c}[\phi,\psi]$. Here, $x=(\tau,\bx)^T$ is a three-component
space-time argument and $S[\psi]=\int\!d^3 x \;\psi^{a \dagger
  }\left[\partial_\tau+iv_{\rm F}(s_1^{a}
  \gamma^{-1}\sigma_1\partial_1+s_2^{a}\sigma_2\partial_2)\right]\psi^a$
the Dirac action of the unperturbed QPs where $v_{\rm F}$ is the Fermi
velocity, $\gamma\equiv t/|\Delta_0|$, $t$ the tight binding energy,
$s_j^a$ are sign factors depending on the node index, and $\sigma_j$
Pauli matrices in particle/hole space~\cite{ASZ}.  Further,
\begin{equation} 
    \label{eq:3} 
    \!S[\phi] = \int \!d^3 x \left[{1\over 2}\Big((\partial_\tau 
\phi)^2\!+\!c^2(\partial_\bx \phi)^2 
       + r \phi^2\Big) + {u\over 4!} \phi^4\right] 
\end{equation} 
is a $\phi^4$-type action controlling the fluctuation behaviour of the
order parameter field. ($c$, $r$, and $u$ are phenomenological
constants, where $r=0$ marks the position of the quantum critical
point into the $d_{x^2-y^2}+is/d_{xy}$ phase.)  Finally, the coupling
between the QP fields and the collective mode $\phi$ is mediated by
the scattering vertex
\begin{equation} 
    \label{eq:4} 
    S_{\rm c}[\phi,\psi] =   \lambda\int \!d^3 x \;   \psi^{a \dagger}  
    \Gamma^a 
    \sigma_3 \psi^a\, 
   \phi, 
\end{equation} 
where $\lambda$ is a coupling constant, and $\Gamma^a$ a symmetry
factor discriminating between the two different types of order
parameters: $\Gamma^a=1$ ($\Gamma^a = (-)^a$) for the case of $is$
($id$) symmetry.  Perhaps the most direct way to understand the
consequences of this coupling mechanism is to integrate over the QP
fields.  This leads to an effective $\phi$-action $S_{\rm
  eff}[\phi]=S[\phi]+ \delta S[\phi]$, where $\delta S[\phi] = -\ln
\langle \exp(-S_{\rm c}[\phi,\psi])\rangle$ measures the tendency of
the system to create/annihilate a Cooper pair in response to
fluctuations of the order parameter and $\langle \dots \rangle = \int
{\cal D}(\psi^\dagger,\psi) \exp(-S[\psi^\dagger, \psi]) (\dots)$ is
the functional average over the QP action. For sufficiently weak
coupling, $\delta S[\phi]$ can be approximated by its second order
expansion in $\phi$ (an RPA type approximation), i.e., $\delta S[\phi]
= { \lambda^2\over 2}\int d^3 x\, d^3 x'\; \phi(x)\chi(x,x')\phi(x')$,
where
\begin{eqnarray} 
  \label{eq:10} 
\chi(x,x') &\equiv&  
{\rm tr\,}\left[\Gamma^a\sigma_3{\cal G}^{aa'}(x,x') \Gamma^{a'}\sigma_3 
{\cal G}^{a'a}(x',x)\right] 
\end{eqnarray} 
is the Cooper pair propagator, `tr' stands for a trace over
particle/hole indices, and ${\cal G}^{aa'}(x,x')= \langle \psi^{a}(x)
\psi^{\dagger a'}(x') \rangle$ is the Gorkov Green function.

Evaluating $\chi$ for the clean case (in which ${\cal G}$ becomes
diagonal in momentum and nodal space), one finds
\begin{equation} 
    \label{eq:5} 
   \delta S[\phi] = -\int d^3k \;|\phi_k|^2\, \left[C_1 - C_2 (\omega_m^2+{\bf 
       k}^2)^{1/2}\right], 
\end{equation} 
where $\int d^3k \equiv T\sum_{\omega_m}\int d^2 k$ denotes a
summation over the three-momentum $k\equiv (\omega_m,{\bf k})^T$, and
$\omega_m$ is a bosonic Matsubara frequency. While the constant
contribution $C_1$ can be absorbed into a shifted value of the
parameter $r$, the non-analytic operator $\sim ({\bf k}^2 +
\omega_m^2)^{1/2} \phi^2$ -- arising from the infrared singularity of
the Dirac operator -- dominantly influences the scaling behaviour of
the model, and holds responsible for the anomalously strong QP
relaxation rate.
 
As a corollary we remark that (ii') the structures above are
insensitive to the symmetry of $\hat \Delta_1(t)$.  The difference
between an $s$- and a $d$-scattering amplitude, respectively, is that
the former is constant while the latter changes sign from node to
node. However, being {\it quadratic} in the scattering vertex, the
pair susceptibility operator (and therefore the induced action $\delta
S$) is oblivious to this difference. Further, (iv') the rough
estimate, $C_1 \sim \lambda^2\int_T^{\omega_0} d\epsilon\,
\nu(\epsilon)/\epsilon$ implies that the same linearity of the DoS,
$\nu(\epsilon) \sim |\epsilon|$, that led to the non-analytic
structure of $\delta S$ excludes the formation of a Cooper instability
in the $\phi$-action.  In contrast, coupling $\phi$ to a fictitious QP
species with an extended Fermi surface, $\nu(\epsilon) = \nu_0,$ would
lead to $C_1(T) \sim \lambda^2 \nu_0 \ln(\omega_0/ T)$, i.e., the
formation of a superconductor transition at a critical temperature
${\tilde T_c} \sim \omega_0 \exp(-\tilde r/\nu_0)$, where $\tilde
r=r/\lambda^2$.

How then do these structures change in the presence of disorder?
Before addressing this question in quantitative detail, let us briefly
summarize a number of key elements entering the theory of the weakly
disordered~\cite{suppress-Delta} system: first, the picture developed
in Refs.~\cite{SaC,KP} essentially relies on the linear singularity of
the low-energy DoS. In contrast, disorder scattering leads to a
randomization of spectral structures over scales $\sim \tau^{-1}$ and,
therefore, to a smearing of the low-energy cusp of the Dirac spectrum.
Relatedly, the QP dynamics crosses over from ballistic at
$|\epsilon|>\tau^{-1}$ to {\it diffusive} at $|\epsilon|<\tau^{-1}$.
As relativistic QP dynamics is an essential input to the
theory~\cite{SaC,KP} one may expect $|\epsilon|\sim \tau^{-1}$ to be a
lower bound for its applicability.  Finally, and in a way to be
discussed in more detail below, the relaxation time (as well as DoS,
spin- and thermal conductance) are affected by mechanisms of quantum
interference, similar to the weak localization corrections known from
the physics of normal metals.

To quantitatively explore the ramification of these mechanisms in the
present context~\cite{ff,KYG,classC}, we consider the configurational
average of the susceptibility operator $\langle \chi(x,x')
\rangle_{\rm dis}$. The most basic effect of impurity scattering will
be that the two Green functions appearing in (\ref{eq:10}) acquire a
self energy, ${\cal G}^{-1}(i\epsilon_n) \to {\cal
  G}^{-1}(i\epsilon_n\zeta ) + i\kappa_{\rm el} \ {\rm
  sgn}(\epsilon_n)$, where the real constants $\kappa_{\rm el}
=(2\tau)^{-1} $ and $\zeta$ are obtained for $|\epsilon| < \tau^{-1}$
from (depending on the microscopic realization of the disorder)
self-consistent Born or $T$-matrix perturbation theory~\cite{leeYA}.

\begin{figure} 
  \medskip
  \centerline{\includegraphics[scale=.225]{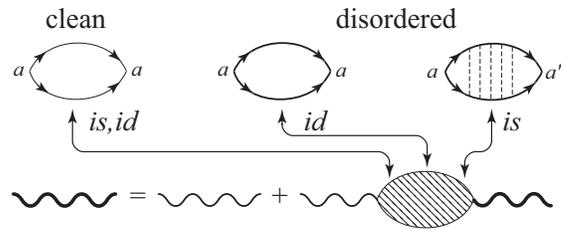}}
\caption{\label{fig1}Top: structure of the polarization operator in 
  the clean, disordered $id$, and disordered $is$ case, respectively.
  Bottom: effective order parameter propagator.}  \vspace*{-0.5cm}
\end{figure}

We next observe that in the \underline{$is$-case} the inclusion of a
Green function self energy is partially compensated for by vertex
corrections.  Evaluating the quasiparticle/order parameter vertex
within a ladder approximation shown in Fig.~\ref{fig1} (the latter
stabilized by the parameter $\gamma\gg 1$), one finds that the vertex
is coupled to a mode $ {\cal D}_{nn'}({\bf q}) = \left(D{\bf q}^2 +
  |\epsilon_n|+|\epsilon_{n'}|\right)^{-1} $ similar to the singular
`diffuson' and `Cooperon' modes of disordered metals. Here,
$D=(\gamma+\gamma^{-1})/(\pi^2\nu_0)$ is the diffusion constant of the
$d$-wave superconductor and $\epsilon_{n,n'}$ are fermionic Matsubara
frequencies. Including this mode in the calculation of the
polarization operator, we find that
\begin{equation} 
\delta S^{is}[\phi]= -\int d^3 k\; |\phi_k|^2\,
\left[ C_1(T) + \dots\,\right],
\label{Sis} 
\end{equation}
where, apart from a $T$-independent shift $\delta r\,(<\!r)$, the
constant term $C_1(T) = \lambda^2\nu_0\ln(\omega_0/T) + \delta r$ now
displays the characteristic low-temperature profile of $s$-wave
superconductors, and the ellipses stand for operators involving
derivatives.  The low-$T$ singularity of $C_1$ implies that at the
critical temperature (\ref{eq:8}) the $\phi$-action becomes unstable,
and a BCS transition into a phase with finite expectation value of the
order parameter ($\langle\phi \rangle\not=0$) and gapped QP states
occurs.  The only significant difference to the BCS transition in
conventional disordered metals is that presently the low-energy DoS
$\nu_0$ -- an essential ingredient to the buildup of a BCS instability
-- is generated by impurity scattering (rather than by the existence
of an extended Fermi surface).  I.e., unlike in normal metals, the
critical temperature is exponentially sensitive to the disorder
concentration.
 
The key phenomenon prescribing much of the physics of the
complementary \underline{$id$-case} is that here the susceptibility
operator is {\it not} affected by vertex corrections. Formally, this
follows from the fact that the mode ${\cal D}$ obtained by summing an
impurity ladder (cf.~Fig.~\ref{fig1}) is isotropic in node space. As a
consequence, the two nodal summations over the symmetry factors
$\Gamma^a = (-)^a$ at the vertices decouple and annihilate all
contributions to the susceptibility operator, safe for the `bare
bubble' without vertex corrections. Heuristically, this mechanism can
be understood by noting that multiple impurity scattering occurring in
transit between two consecutive interactions with the order parameter
field distributes the QP amplitude homogeneously over all nodes. The
vanishing of the nodal average of $\hat\Delta^{d_{xy}}$ then implies
the absence of multiple scattering contributions to the pair
propagator.
 
Evaluating Eq.~(\ref{eq:10}) for the self-energy decorated Green
functions, we find
\begin{equation} 
\delta S^{id}[\phi] = -\int d^3 k\;|\phi_k|^2 \,\left[C_1 - C_2 
\left(|\omega_m| 
+ Dk^2\right)\right], 
\label{Sid}
\end{equation}
with the (now again finite at $T\to0$) constant $C_1 \sim
\lambda^2\omega_D f(\omega_D{\tau}\zeta)$, where $f(x)$ decreases from
1 at $x\gg 1$ to $x$ at $x\to 0$, and $C_2 \sim \lambda^2$.
Importantly, a non-analytic contribution $\sim |\omega_m| \phi^2$
survives the blurring of the low-energy spectrum by the imaginary part
of the self-energy.  The continued presence of an operator of
engineering dimension $1$ indicates that the collective mode displays
fluctuation behaviour similar to that in the clean case.  Relatedly, a
straightforward calculation along the lines of Ref.~\cite{KP} shows
that in the vicinity of the instability ($r\ll C_2|\epsilon|$)
$\kappa_{\rm in} \sim |\epsilon|$, as in the absence of disorder,
while for $r\gg C_2|\epsilon|$ we obtain Fermi liquid behaviour
$\kappa_{\rm in} \sim \epsilon^2$.
 
\begin{figure} 
  \medskip
  \centerline{\hspace{-1.4cm}\includegraphics[scale=0.325]{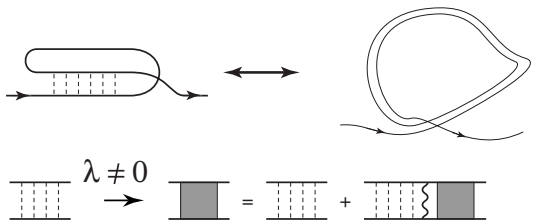}\vspace{-2cm}}
\caption{\label{fig2} Top: quantum interference correction to the 
  self energy (left) and corresponding Feynman path (right). Bottom:
  renormalization of the interference process by interactions.}
\vspace*{-0.5cm}
\end{figure}

Having identified the two terms $\kappa_{\rm in}$ and $\kappa_{\rm
  el}$, we proceed to discuss the contribution of disorder-induced
quantum interference to the QP self energy. To leading order in an
expansion in the small parameter $g^{-1}$, where $g= D\nu_0$, the soft
modes ${\cal D}$ couple to the self-energy operator through the
diagram shown in Fig.~\ref{fig2}. Physically (cf.~Fig.~\ref{fig2}),
this diagram describes the self-interference of a QP traversing a
scattering path twice.  In close analogy to the weak localization
corrections to the conductance of normal metals, these processes lead
to a negative logarithmic correction $\kappa_{\rm sing}(\epsilon) \sim
g^{-1} \ln(\tau|\epsilon|)$ to the self energy. In passing we note
that the singularity of the interference correction to the DoS entails
the existence of a crossover scale $\omega^\ast \sim \tau^{-1}
\exp(-g)$, below which a proliferation of diffusion modes holds
responsible for both, Anderson localization of the QP states and a
linear vanishing of their DoS, $\nu(\epsilon) \sim \nu_0
|\epsilon|/\omega^\ast$~\cite{ASZ,s-note}.

While the logarithmic singularity of the self energy is primarily
caused by disorder scattering, the presence of a dynamical $id$ order
parameter necessitates to replace the bare diffusion mode by the
`interacting' mode shown in Fig.~\ref{fig2} (bottom).  However, the
quantitative summation over these self-interaction processes merely
leads to an inessential renormalization of the prefactor, $g^{-1}\to
g^{-1}(1+F)$, where $F\ll 1$ is an interaction-dependent constant,
playing a role similar to the Fermi-liquid constants in normal metals.
 
Similar logarithmic corrections, weakly renormalized by QP
interactions, are found for the (spin-) conductance and the
DoS~\cite{next,gauge}.  These findings conform with the RG analysis of
Ref.~\cite{classC} which showed (albeit for the case of globally
broken time-reversal invariance -- symmetry class $C$ in the notation
of Ref.~\cite{classes}, while the present problem falls into the more
complex symmetry class $C$I) that interactions only weakly affect the
low-energy disorder-generated interference phenomena~\cite{is-note}.

Summarizing, we have explored the interplay of disorder scattering and
critical fluctuations of an $is$- or $id$-order parameter component in
two-dimensional cuprate superconductors. In the $is$-case, the most
prominent effect caused by disorder is a secondary BCS transition
whose critical temperature sensitively depends on the impurity
concentration. In the complementary $id$-case, the addition of
disorder to the system leaves much of the phenomenology derived in
Refs.~\cite{SaC,KP} intact.  Notably, for temperatures $T<|\epsilon|$,
the characteristic singularity $\kappa_{\rm in}\sim |\epsilon|$
pertains to the disordered case.  Heuristically, these phenomena can
be understood by noting that in the presence of disorder (and for mass
parameters fine tuned to criticality) the collective mode
corresponding to the order parameter amplitude {\it diffuses} and,
therefore, continues to act as a strong agent of QP scattering.
Finally, we included the (essentially disorder-generated) formation of
logarithmic singularities at low energies into our discussion of the
relaxation rate. Given that even moderate impurity
concentrations~\cite{ong} give rise to scattering times comparable to
the energies (${\cal O}(20{\rm K})$) estimated~\cite{SaC} as
characteristic for the fluctuations of $\Delta_1$, we believe that,
for sufficiently low temperatures, our findings may be made visible
experimentally.
 
\begin{acknowledgments} 
 We are grateful to
%%%%%%%%%%%%%%%%%%%%%%%%%%%%%%%% 
 D.V. Khveshchenko, A.W.W. Ludwig,  A.D. Mirlin,
 J. Paaske, S. Sachdev, M. Vojta and A.G. Yashenkin 
 %%%%%%%%%%%%%%%%%%%%%%%%%%%%%%%% 
for valuable discussions. 
 %%%%%%%%%%%%%%%%%%%%%%%%%%%%%%%% 
I.V.G. was supported by the Schwerpunktprogramm ``Quanten-Hall-Systeme'',
the SFB195 der Deutschen Forschungsgemeinschaft, and by the RFBR.
%%%%%%%%%%%%%%%%%%%%%%%%%%%%%%%%  
\end{acknowledgments}


\begin{thebibliography}{}
\bibitem{suppress-Delta} We assume $\Delta_0$ to be so large that the
  pair breaking effect of disorder can be neglected.  Furthermore, we
  do not consider intrinsic randomness in the parameter
  $r$~\cite{book} as it does not qualitatively change our results.
\bibitem{s-note} In the $is$-case, for $\omega^\ast\sim \tilde T_c$,
  the BCS instability may interfere with the formation of strong
  localization phenomena. As $\omega^\ast$ and $\tilde T_c$ depend on
  fundamentally different input parameters, either case $\omega^\ast <
  \tilde T_c$ or $\omega^\ast> \tilde T_c$ is conceivable~\cite{next}.
\bibitem{is-note} While these statements apply for a generic form of
  the interaction ($r\not=0$), fine tuning the collective mode into a
  critical regime will lead to more pronounced interaction-generated
  phenomena~\cite{next} in both, $is$ and $id$, cases.
\end{thebibliography}
\end{document}